\documentclass[a4paper,pre,aps,twocolumn,superscriptaddress,showpacs,floatfix]{revtex4-1}

\usepackage[utf8]{inputenc}
\usepackage{amssymb}
\usepackage{latexsym}
\usepackage{amsmath}

\usepackage{natbib}
\usepackage[dvips]{graphicx}

\usepackage{longtable} 
\usepackage{rotating}

\newcommand{\Lg}{\Lambda}
\newcommand{\ag}{\alpha}
\newcommand{\bg}{\beta}
\newcommand{\eA}{\epsilon}
\newcommand{\lc}{\left [}
\newcommand{\rc}{\right ]} 
\newcommand{\lp}{\left (}
\newcommand{\rp}{\right )}

\begin{document}

\bibliographystyle{elsarticle-harv}

\title{Spreading of intolerance under economic stress: results from a model
with reputation.}

\author{Luis A. Martinez-Vaquero}
\affiliation{Grupo Interdisciplinar de Sistemas Complejos (GISC),
Departamento de Matem\'aticas, Universidad Carlos III de Madrid,
28911 Legan\'es, Madrid, Spain}
\affiliation{Current address: Artificial Intelligence Lab, Vrije Universiteit
Brussel, Pleinlaan 2, 1050 Brussels, Belgium}

\author{Jos\'e A. Cuesta}
\affiliation{Grupo Interdisciplinar de Sistemas Complejos (GISC),
Departamento de Matem\'aticas, Universidad Carlos III de Madrid,
28911 Legan\'es, Madrid, Spain}
\affiliation{Instituto de Biocomputaci\'on y F\'\i sica de Sistemas
Complejos (BIFI), Universidad de Zaragoza, 50009 Zaragoza, Spain}

\begin{abstract}

When a population is engaged in successive prisoner's dilemmas, indirect
reciprocity through reputation fosters cooperation through the emergence of
moral and action rules. A simplified model has recently been proposed where
individuals choose between helping or not others, and are judged good or bad
for it by the rest of the population. The reputation so acquired will condition
future actions. In this model, eight strategies (referred to as `leading
eight') enforce a high level of cooperation, generate high payoffs and are
therefore resistant to invasions by other strategies. Here we show that, by
assigning each individual one out of two labels that peers can distinguish
(e.g., political ideas, religion, skin colour...) and allowing moral and action
rules to depend on the label, intolerant behaviours can emerge within
minorities under sufficient economic stress. We analyse the sets of conditions
where this can happen and also discuss the circumstances under which tolerance
can be restored. Our results agree with empirical observations that correlate
intolerance and economic stress, and predict a correlation between the degree
of tolerance of a population and its composition and ethical stance.

\end{abstract}



\maketitle

\section{Introduction}

Different kinds of discriminatory behaviour based on membership to different
groups---identified, for example, by visible tags \cite{hamilton:1964b}---has
been reported both in human \cite{sumner:1906, levine:1972, Hirschfeld:1996,
bernhard:2006} and in animal societies \cite{haig:1996, keller:1998,
Queller:2003, Summers:2005, Sinervo:2006}.  They can be broadly classified into
two main types: in-group favoritism and out-group hostility \cite{ray:1986,
struch:1989, cashdan:2001}. Recent studies of in-group and out-group
cooperation show that in-group favouritism can emerge in the framework of
indirect reciprocity \cite{masuda:2012, nakamura:2012, oishi:2012}.

The concept of indirect reciprocity \cite{sugden:1986, alexander:1987} has
been introduced to explain the emergence of cooperation in society, where many
interactions between the same individuals have low chances to be repeated.
Contrary to direct reciprocity \cite{trivers:1971}, indirect reciprocity
implies that individuals receive the consequences of their actions, not
directly from their opponents, but indirectly through society. Indirect
reciprocity---and its related reputation concept---has proven to be an
important mechanism for the emergence and sustainment of cooperation in
small-scale human \cite{dufwenberg:2001, milinski:2002, panchanathan:2004,
semmann:2004, suzuki:2007} and non-human \cite{bshary:2006} societies. It also
plays an important role in communication networks \cite{bolton:2004,
keser:2002}. In these reputation-based models, individuals have an opinion
about every interaction they witness and assign a reputation to the individuals
involved accordingly \cite{nowak:1998, ohtsuki:2004, brandt:2004, nowak:2005,
milinski:2002}. The nature of their future interactions with those individuals
will be determined by the reputation they have assigned to them.

In a stylised model of indirect reciprocity Ohtsuki and Iwasa
\cite{ohtsuki:2004} and Brandt and Sigmund \cite{brandt:2004} classified the
strategies involved in these games attending to action and assessment modules.
The action module prescribes what to do---whether cooperate or not---given the
reputations of the individuals involved. The assessment module assigns
reputation to the interacting individuals according to a moral code. To do it
three elements can be judged: the nature of the interaction, the reputation of
the recipient, and the reputation of the donor.  Strategies are thus classified
as first, second, or third order depending on whether the first, the first and
the second, or all the three elements are taken into account to assign a
reputation to the donor. Ohtsuki and Iwasa \cite{ohtsuki:2004} studied the
evolutionarily stability of third order strategies which share the same moral
assessment module, and later Martinez-Vaquero and Cuesta \cite{martinez:2013}
extended the study confronting strategies with different moral codes. Among the
evolutionary stable strategies (ESS) found, there are eight strategies---the
so-called \textit{Leading Eight}---that are considerably more efficient (get
very high payoff) and coherent (in terms of consistency between their moral and
action modules) than the rest of them. All these strategies foster cooperation
through indirect reciprocity.

This simple model of indirect reciprocity can be readily extended to study the
emergence of discriminatory behaviours by introducing different groups of
individuals identified by external signs (these may include physical or
cultural traits). Our goal in this work is to analyse under which conditions
an intolerant strategy (understood as out-group hostility) can invade a
tolerant population that follows one of the Leading Eight strategies. Whenever
intolerance spreads, we are also interested in whether tolerance can be restored
by introducing some kind of external incentives. Our hope is that this study
sheds some light into the causes for the emergence of intolerance in societies
and the mechanisms to alleviate this social burden.

The paper is structured as follows. In section \ref{sec:model} we introduce the
model, which is mathematically implemented in section \ref{sec:implementation}.
Our results are shown in section \ref{sec:results} and discussed in section
\ref{sec:discussion}.

\section{Model}
\label{sec:model}

Our model is an extension of that used in a previous work
\cite{martinez:2013}---itself a modification of the indirect reciprocity model
based on reputation introduced by Brandt and Sigmund \cite{brandt:2004} and
later investigated by Ohtsuki and Iwasa \cite{ohtsuki:2004,ohtsuki:2006b}.

We consider an infinite, well-mixed population, where every individual is aware
of every action performed and produces a moral judgment---which leads to
a reputation assignment---on it. Each time step a pair of individuals is
randomly drawn---with equal probability---from the population. One of them 
plays as \emph{donor} and the other one as \emph{recipient}. The donor can
decide to pay a cost $c>0$ to help (C) the recipient or not (D). Any helped
recipient gains a payoff $b>c$; otherwise there is no payoff whatsoever.

This \emph{action} is observed by every individual in the population (including
themselves). Thus everyone makes a private judgment of the donor for this
action. The observer's \emph{moral assessment} will decide the
reputation---good (G) or bad (B)---she assigns to the donor. Accordingly,
everybody has a private opinion of every other individual---even of herself.

By repeating this process the population eventually reaches a steady state.
The average payoff of this repeated game is then computed for every individual.
By assuming a very large population we effectively neglect direct
reciprocity---the probability that two individuals meet again is very small.

Strategies are defined by two modules: the action rules, which tells the donor
how to interact with the recipient, and the moral rules, which prescribe a
reputation for the donor of every witnessed action. As in our previous paper
\cite{martinez:2013}, we will consider third order strategies.

The action rules determine what the donor must do (either help or refuse to
help) given the reputation of both players. Specifically, $a_{\alpha\beta}=1$
(C) if a strategist with reputation $\alpha$ helps an individual with
reputation $\beta$ and $0$ (D) otherwise.

The moral assessments tell the individual if the action just witnessed should
be judged good or bad, hence revising the donor's reputation. Specifically,
$m_{\alpha\beta}(a)=1$ (G) if the observer assigns good reputation to a
donor she previously judged $\alpha$, who performs an action $a$ on an
recipient she previously judged $\beta$, and is $0$ (B) otherwise.

Thus each strategy is defined by 12 numbers: 4 for the action module and 8 for
the moral module. This amounts to 4096 different strategies in total.

Besides, we will assume that players sometimes make errors when trying to help
another individual \cite{leimar:2001, ohtsuki:2004, panchanathan:2003,
fishman:2003, lotem:1999}. Thus, with a probability $\eA$ a donor defects
regardless of her action rules and with $1-\eA$ she performs the action she
planned to. One possible source of this error---which will be important in
interpreting the results of this work---is scarcity of resources. Despite her
willingness to help, an individual may fail to do it because she cannot
afford the cost $c$. Errors due to misjudgements are excluded (see
\cite{martinez:2013} for a justification).

Discrimination requires at least two different subpopulations that can be
clearly identified by everybody. Thus we divide our population into two
classes, A and B. The fraction of A individuals will be denoted $y$. A
\textit{tolerant} player will act disregarding the individuals' class when
deciding her action or judgement. An \textit{intolerant} individual acts and
judges like a tolerant one with the following exceptions:
\begin{enumerate} 
\item Individuals of the other class are never helped.
\item Donors of the other class are judged bad under any circumstance.
\item Individuals of her own class who do not help individuals of the other
class are always judged good.
\item Individuals of her own class who do help individuals of the other class
are always judged bad.
\end{enumerate}

\section{Implementation}
\label{sec:implementation}

Our goal is to determine, with the assumptions of the model, under which
conditions a small population of intolerant (tolerant) mutants $M$ can invade a
tolerant (intolerant) resident population $R$ made of A-type and B-type
individuals in fractions $y$ and $1-y$ respectively. This fraction $y$ will be
kept constant and is therefore a parameter of the model. We will only consider
scenarios where all individuals of the same class in $R$ show the same kind of
behavior---either tolerant or intolerant---toward the other class. Likewise,
we will limit ourselves to determine when the resident population can resist
the invasion of mutants. Determining the final composition of the population
if the invasion succeeds is a more complex problem that we will tackle in this
paper.

Without loss of generality, in the following calculations mutants will be
assumed to belong to class A. Likewise, we will only need to consider whether
A-residents imitate the mutant or not. If a B-resident imitates the A-mutant
(its tolerant or intolerant character) this can be treated as an invasion of
B-residents by a B-mutant separately---the reason being that mutants are
present in so small a fraction that the probability that two mutants interact
is negligible.

Accordingly the different invasion scenarios will be denoted
$t_M|t_{\text{A}}t_{\text{B}}$, where $t_X$ can be either T (tolerant
individuals) or I (intolerant individuals), $M$ denotes A-type mutants, and A
and B the two subpopulations. The four possible scenarios are:
\begin{description}\setlength{\itemsep}{-2pt}
\item[\rm I$|$TT] intolerant A-mutants try to invade a tolerant population;
\item[\rm T$|$IT] tolerant A-mutants try to invade a population made of
intolerant A-residents and tolerant B-residents;
\item[\rm I$|$TI] intolerant A-mutants try to invade a population made of
tolerant A-residents and intolerant B-residents;
\item[\rm T$|$II] tolerant A-mutants try to invade an intolerant population.
\end{description}

The condition for A-residents to resist the invasion of A-mutants is
$W(A|R)>W(M|R)$, where $W(A|R)$ and $W(M|R)$ are the average payoffs received
by A-type residents and by mutants, respectively. They can be obtained as
\begin{equation}
\begin{split}
W(A|R)&=b\ \Theta_{R,A} -c\ \Theta_{A,R}, \\
W(M|R)&=b\ \Theta_{R,M} -c\ \Theta_{M,R}, \\
\Theta_{R,A}&=y\ \theta_{A,A}+(1-y)\ \theta_{B,A}, \\
\Theta_{A,R}&=y\ \theta_{A,A}+(1-y)\ \theta_{A,B}, \\
\Theta_{R,M}&=y\ \theta_{A,M}+(1-y)\ \theta_{B,M}, \\
\Theta_{M,R}&=y\ \theta_{M,A}+(1-y)\ \theta_{M,B},
\end{split}
\label{eq:WA1_gen}
\end{equation}
where $\theta_{i,j}$ is the probability that an $i$-strategist helps a
$j$-strategist. If $i$-strategists are intolerant and $j$-strategists belong to
a different class $\theta_{i,j}=0$; otherwise these probabilities can be
obtained as
\begin{equation}
\theta_{i,j}=(1-\eA) \sum_{\ag\bg} \chi_\ag(g_i^i) \chi_\bg(g_j^i)
a_{\ag\bg},
\label{eq:thetaij}
\end{equation}
where $g_i^j$ is the fraction of $i$-strategists that are considered good by
the $j$-strategists. We have introduced the auxiliary function
\begin{equation}
\chi_{\gamma}(x)=\gamma x+(1-\gamma)(1-x).
\label{eq:chi}
\end{equation}
Thus if $x$ represent the fraction of good individuals, then $\chi_1(x)=x$
and $\chi_0(x)=1-x$ are the fraction of `good' and `bad' individuals,
respectively.

In order to calculate the fractions $g_i^j$ we first need to compute the
fractions $x_i^{\Lg_A \Lg_B \Lg_M}$ of $i$-individuals with reputations
$\Lg_A$, $\Lg_B$, and $\Lg_M$ according to A-residents, B-residents, and
mutants, respectively. Here $\Lg_i$ can be G (for good reputation), B (for
bad reputation), or * (for any reputation, G or B). For instance, $x_i^{G*B}$
represents the fraction of $i$-type individuals who are considered good by
A-residents and bad by mutants, regardless of B-residents' opinion. Hence
$x_i^{G*B}=\sum_{\Lg_B} x_i^{G \Lg_B B}$. Thus
\begin{equation}
g_{i}^{A} = x_i^{G**}, \qquad
g_{i}^{B} = x_i^{*G*}, \qquad
g_{i}^{M} = x_i^{**G}.
\label{eq:g}
\end{equation}

It turns out that not all $x_i^{\Lg_A \Lg_B \Lg_M}$ are necessary to compute
$g_i^j$ (see Appendix for details). In general, all we need is to obtain
$x_A^{\Lg_A*\Lg_M}$, $x_M^{\Lg_A*\Lg_M}$, and for some scenarios also
$x_B^{\Lg_A*\Lg_M}$ or $x_B^{*\Lg_B\Lg_M}$. Since we are considering an
infinite population, self-interactions are negligible. Under the assumption
that the rate of mutants is very low, donors will only interact with residents
and therefore the dynamics of these fractions will be given by
\begin{align}
\frac{dx_{A}^{\Lg_A*\Lg_M}}{dt} &=y\ T_{A,A}^{\Lg_A\Lg_M} +
     (1-y)\ F_{A,B}^{\Lg_A\Lg_M} - x_A^{\Lg_A*\Lg_M}, \label{eq:evol_xA_gen} \\
\frac{dx_{B}^{\Lg_A\Lg_B*}}{dt} &=(1-y)\ T_{B,B}^{\Lg_A\Lg_B} +
     y\ F_{B,A}^{\Lg_A\Lg_B} - x_B^{\Lg_A*\Lg_B},  \label{eq:evol_xB1_gen} \\
\frac{dx_{B}^{*\Lg_B\Lg_M}}{dt} &=(1-y)\ T_{B,B}^{\Lg_B\Lg_M} +
     y\ F_{B,A}^{\Lg_B\Lg_M} - x_B^{*\Lg_B\Lg_M}, \label{eq:evol_xB2_gen} \\
\frac{dx_{M}^{\Lg_A*\Lg_M}}{dt} &=y\ T_{M,A}^{\Lg_A\Lg_M} +
     (1-y)\ F_{M,B}^{\Lg_A\Lg_M} - x_M^{\Lg_A*\Lg_M}, \label{eq:evol_xM_gen}
\end{align}
where the interactions with recipients of the same and of the opposite class
have been split into $T_{i,j}^{\Lg_l\Lg_m}$ and $F_{i,j}^{\Lg_l\Lg_m}$,
respectively. Fractions $F_{i,j}^{\Lg_l\Lg_m}$ depend on the scenario we are
considering whereas $T_{i,j}^{\Lg_l\Lg_m}$ are common for all of them and can
be obtained as
\begin{align}
T_{i,A}^{\Lg_A\Lg_M} =& \sum_{\ag_A\ag_M\bg_A\bg_M} x_i^{\ag_A*\ag_M} x_A^{\bg_A*\bg_M}
\nonumber \\
&\times R^{\Lg_A\Lg_M}(\ag_A\bg_A,\ag_M\bg_M|\ag_i\bg_i) \\
T_{B,B}^{\Lg_A\Lg_B} =& \sum_{\ag_A\ag_B\bg_A\bg_B} x_B^{\ag_A\ag_B*} x_B^{\bg_A\bg_B*}
\nonumber \\
&\times R^{\Lg_A\Lg_B}(\ag_A\bg_A,\ag_B\bg_B|\ag_B\bg_B) \\
T_{B,B}^{\Lg_B\Lg_M} =& \sum_{\ag_B\ag_M\bg_B\bg_M} x_B^{*\ag_B\ag_B} x_B^{*\bg_B\bg_M}
\nonumber \\
&\times R^{\Lg_B\Lg_M}(\ag_B\bg_B,\ag_M\bg_M|\ag_B\bg_B)
\end{align}
where $R^{\Lg_l\Lg_m}(\ag_l\bg_l,\ag_m\bg_m|\ag_i\bg_i)$ is the probability
that an $i$-type donor with reputation $(\ag_l,\ag_m)$ acting on a recipient
with reputation $(\bg_l,\bg_m)$, is assigned a reputation $(\Lg_l,\Lg_m)$.
Reputations are as given by $l$-individuals and $m$-individuals respectively.
This probability is obtained as
\begin{equation}
\begin{split}
R  ^{\Lg_l\Lg_m} & (\ag_l\bg_l,\ag_m\bg_m|\ag_i\bg_i) = \\
&= (1-\eA)\,\delta \big( \Lg_l, m_{\ag_l\bg_l}(a_{\ag_i\bg_i}) \big) 
\delta \big( \Lg_m, m_{\ag_m\bg_m}(a_{\ag_i\bg_i})\big) \\
&+\, \eA\,\delta \big( \Lg_l, m_{\ag_l\bg_l}(D) \big) 
 \delta \big( \Lg_m, m_{\ag_m\bg_m}(D)\big).
\end{split}
\end{equation}

In general, Eqs.~\eqref{eq:evol_xA_gen}--\eqref{eq:evol_xM_gen} need to be
solved numerically. Nevertheless, for the different scenarios that we will
consider, some simplifications can be made. The details are deserved to the
Appendix.

\section{Results}
\label{sec:results}

\begin{table*}[!ht] 
  \begin{center}
    \begin{tabular}{|c|cccccccc|cccc|}
     \hline
	& $m_{GG}(C)$ & $m_{GG}(D)$ & $m_{GB}(C)$ & $m_{GB}(D)$ & $m_{BG}(C)$ &
$m_{BG}(D)$ & $m_{BB}(C)$ & $m_{BB}(D)$ & $a_{GG}$ & $a_{GB}$ & $a_{BG}$ & $a_{BB}$ \\
     \hline
  Ia  & G & B & G & G & G & B & G & B & C & D & C & C  \\
  Ib  & G & B & B & G & G & B & G & B & C & D & C & C   \\
       \hline
  IIa & G & B & G & G & G & B & G & G & C & D & C & D \\
  IIb & G & B & G & G & G & B & B & G & C & D & C & D  \\
  IIc & G & B & B & G & G & B & G & G & C & D & C & D  \\
  IId & G & B & B & G & G & B & B & G & C & D & C & D  \\
       \hline
  IIIa& G & B & G & G & G & B & B & B & C & D & C & D  \\
  IIIb& G & B & B & G & G & B & B & B & C & D & C & D  \\
     \hline  
    \end{tabular}
 \caption{Leading eight strategies. The first eight columns describe the
moral assessment; the last four, the action rules. Strategies are classified
into three groups (I, II, and III; see text).}
  \end{center}
\label{tab:leading8}
\end{table*}

We focus our analysis on the \emph{leading eight} strategies introduced by
Ohtsuki and Iwasa \cite{ohtsuki:2004}. In a previous paper \cite{martinez:2013}
we proved that, for reasonable values of benefit-to-cost and error rates, these
eight strategies are evolutionarily stable against the invasion of any other
strategy. On the other hand, Ohtsuki and Iwasa \cite{ohtsuki:2004} classified
the leading eight in three groups (Table~\ref{tab:leading8}), according to
their stability against invasions by fully defective action rules (AllD).
Neglecting errors in the moral assessment, Groups I and II can be invaded by
AllD provided
\begin{equation}
\frac{b}{c} < 1 + \eA + \eA^2 + \dots = \frac{1}{1-\eA},
\end{equation}
in other words, if the error in the action $\eA> (b-c)/b$. Invading Group III
requires $b<c$ though.

Therefore, for a given $b>c$, if $\eA$ is sufficiently large, any Group I
or Group II strategy can be invaded by defectors, whereas Group III strategies
always resist invasions. As we will show below, these are the same conditions
under which intolerance can spread, but intolerant strategies obtain a higher
payoff than AllD because they are indistinguishable from the strategy used by
residents of their same group. Accordingly, when two distinguishable groups
coexist in a population, intolerant strategies are preferred over AllD.

\begin{figure*}
\begin{center}
\includegraphics[width=6in]{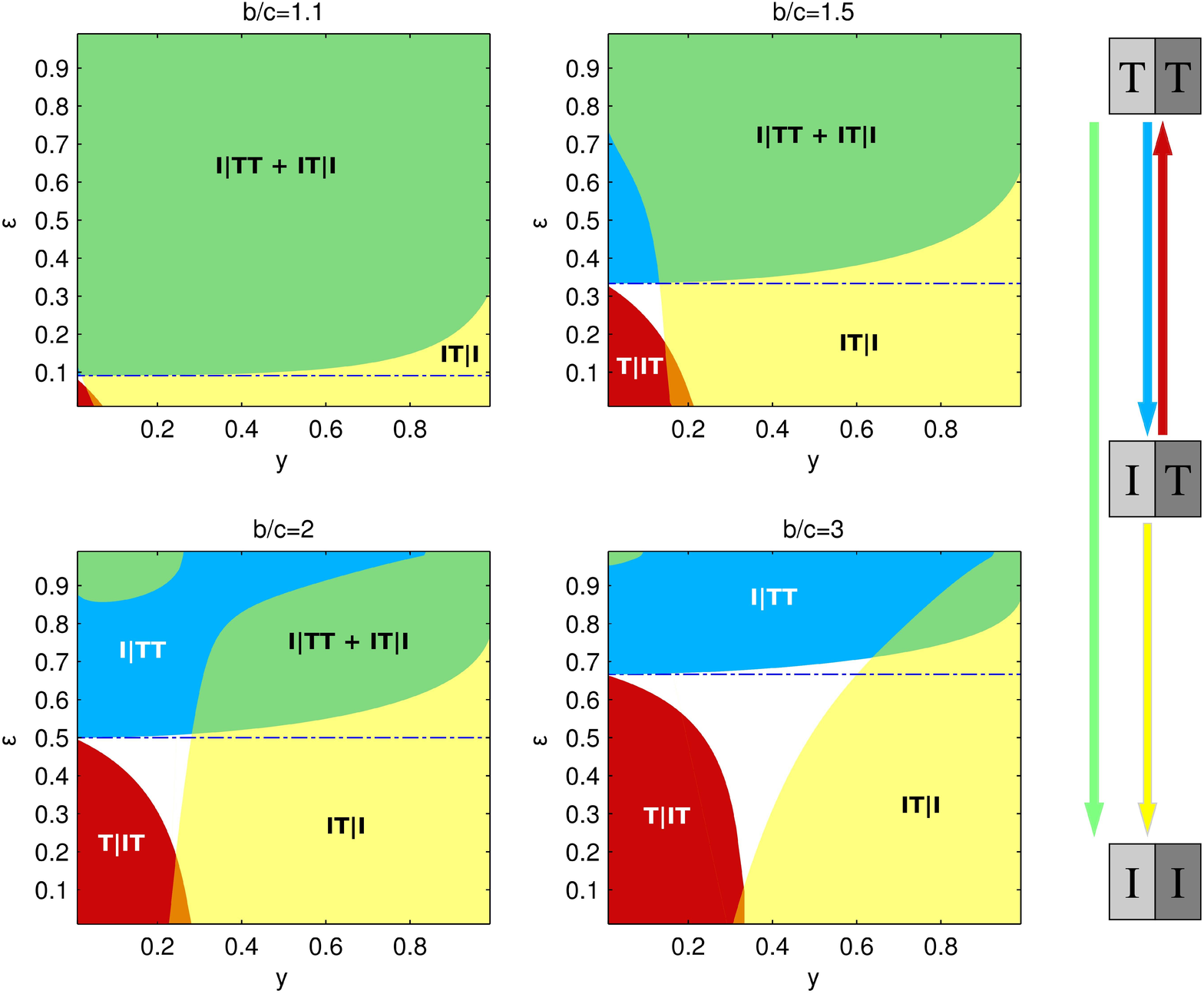}
\end{center}
\caption{
(Color online.)
Values for $\eA$ and $y$ indicating the different regions where invasions
$I|TT$, $T|IT$ and $IT|I$ are successful for strategy Ia.
Above the horizontal line, also AllD can invade a tolerant population.
}
\label{fig:Ia}
\end{figure*}

\begin{figure*}
\begin{center}
\includegraphics[width=6in]{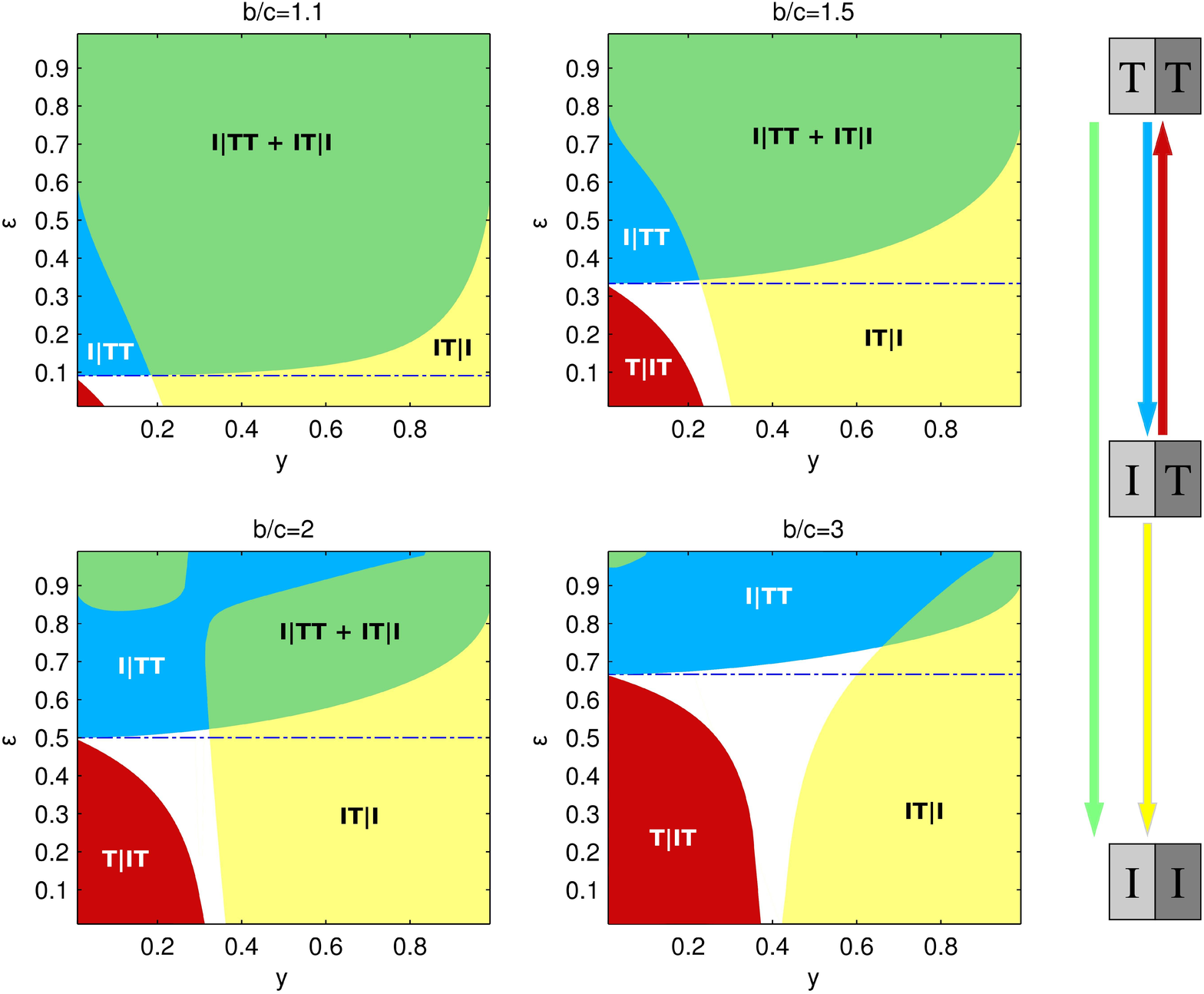}
\end{center}
\caption{
(Color online.)
Same as Fig.~\ref{fig:Ia} but for strategy Ib.
}
\label{fig:Ib}
\end{figure*}

\begin{figure*}
\begin{center}
\includegraphics[width=6in]{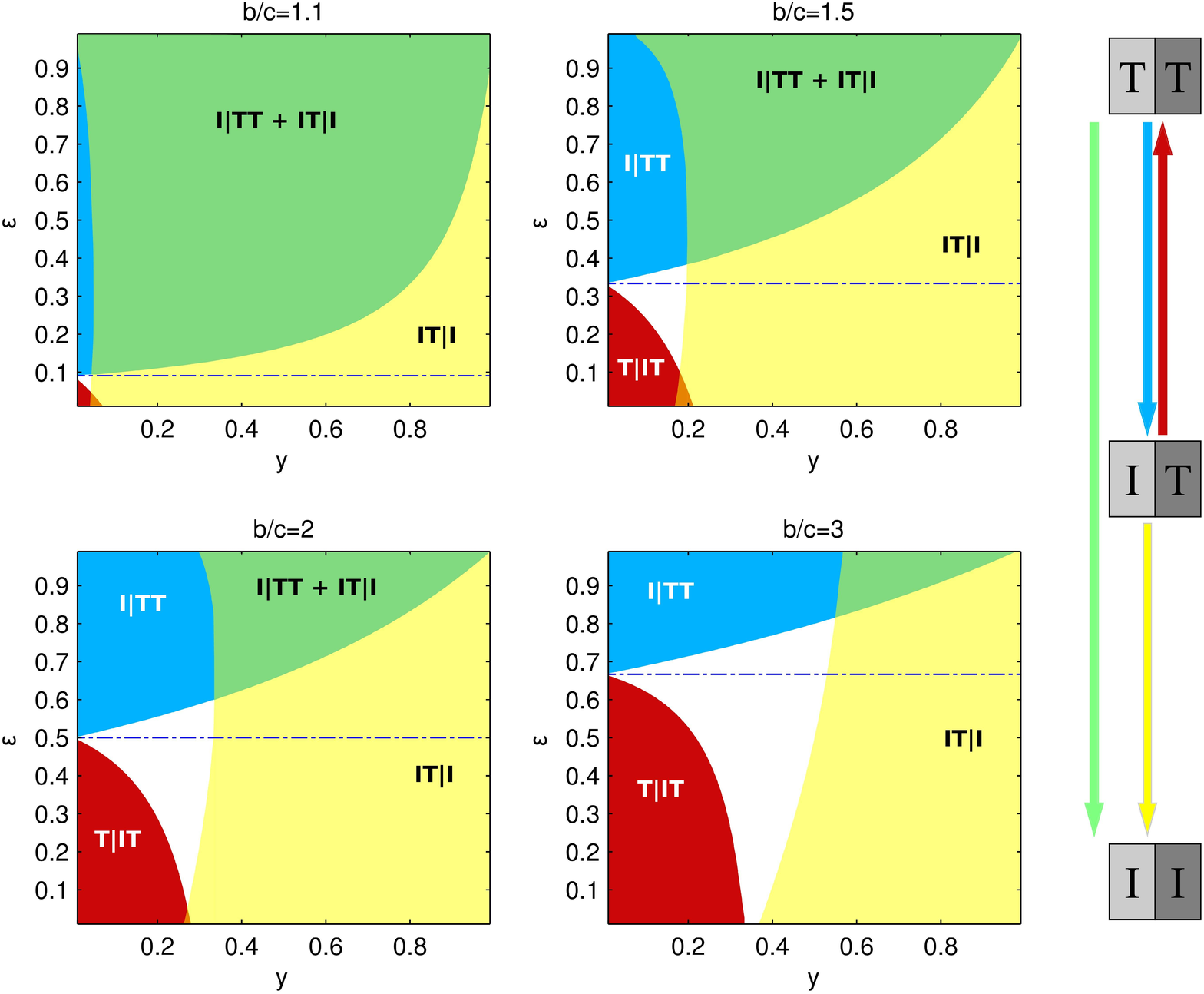}
\end{center}
\caption{
(Color online.)
Same as Fig.~\ref{fig:Ia} but for strategy IIa.
}
\label{fig:IIa}
\end{figure*}

\begin{figure*}
\begin{center}
\includegraphics[width=6in]{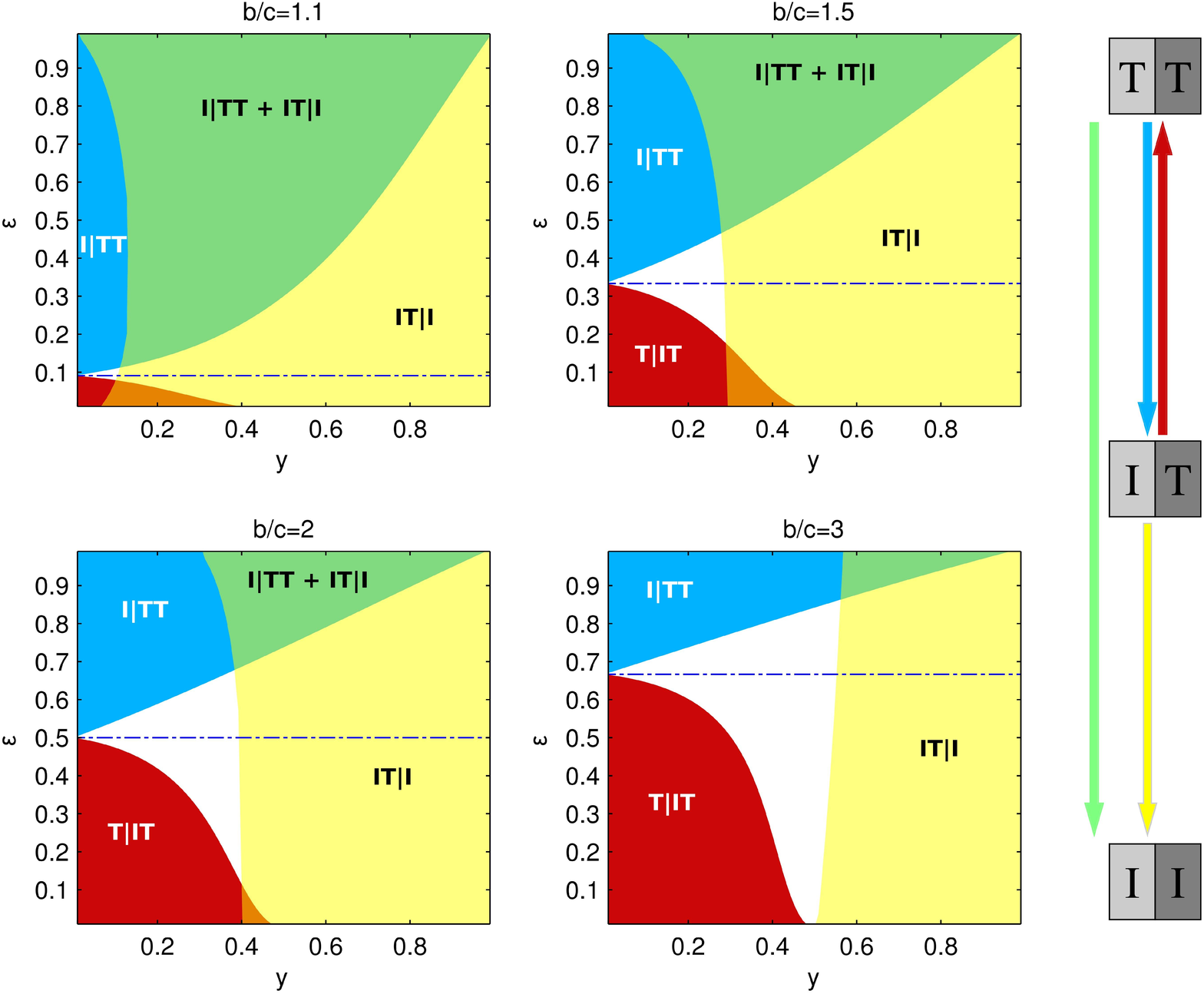}
\end{center}
\caption{
(Color online.)
Same as Fig.~\ref{fig:Ia} but for strategy IIb.
}
\label{fig:IIb}
\end{figure*}

\begin{figure*}
\begin{center}
\includegraphics[width=6in]{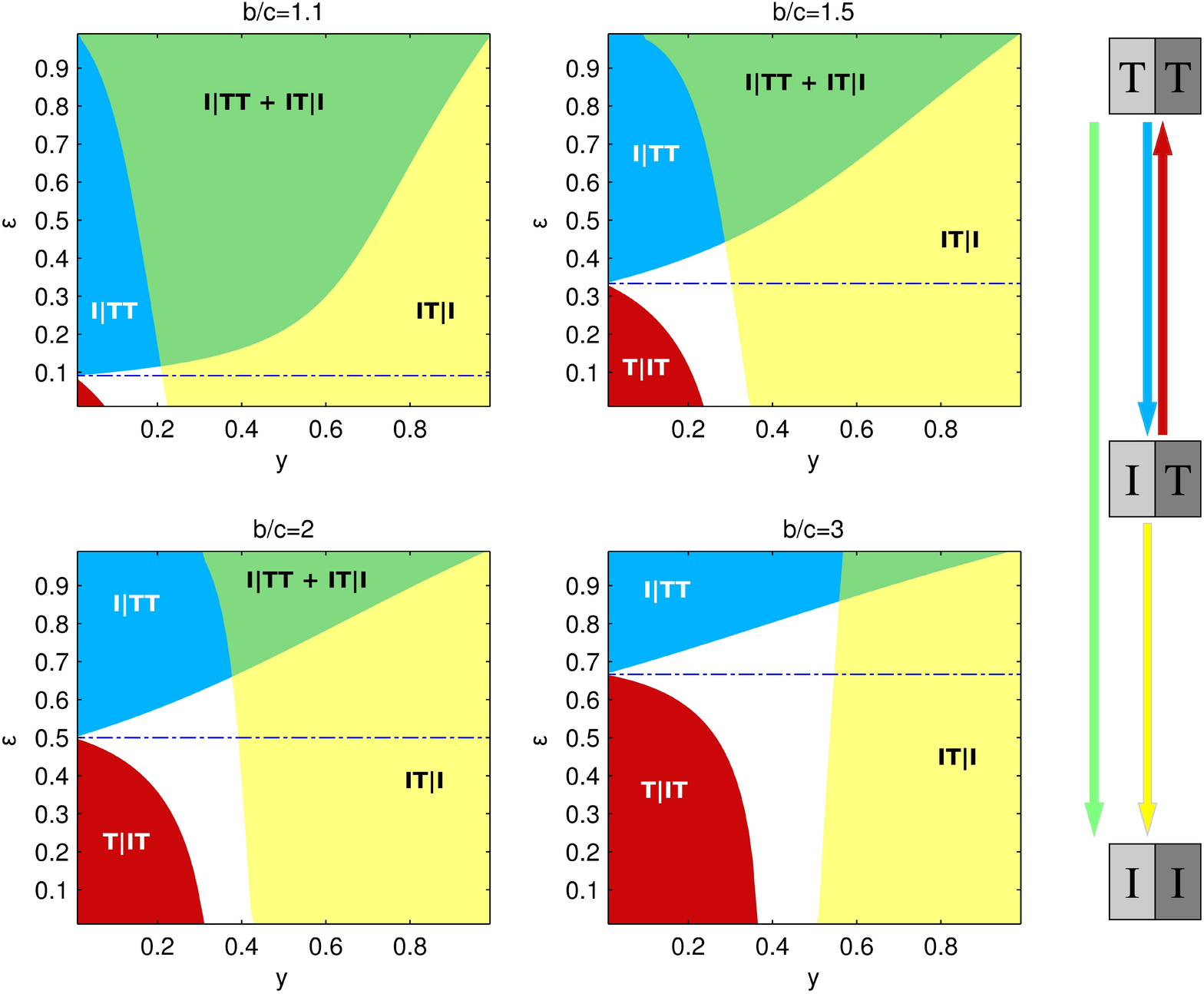}
\end{center}
\caption{
(Color online.)
Same as Fig.~\ref{fig:Ia} but for strategy IIc.
}
\label{fig:IIc}
\end{figure*}

\begin{figure*}
\begin{center}
\includegraphics[width=6in]{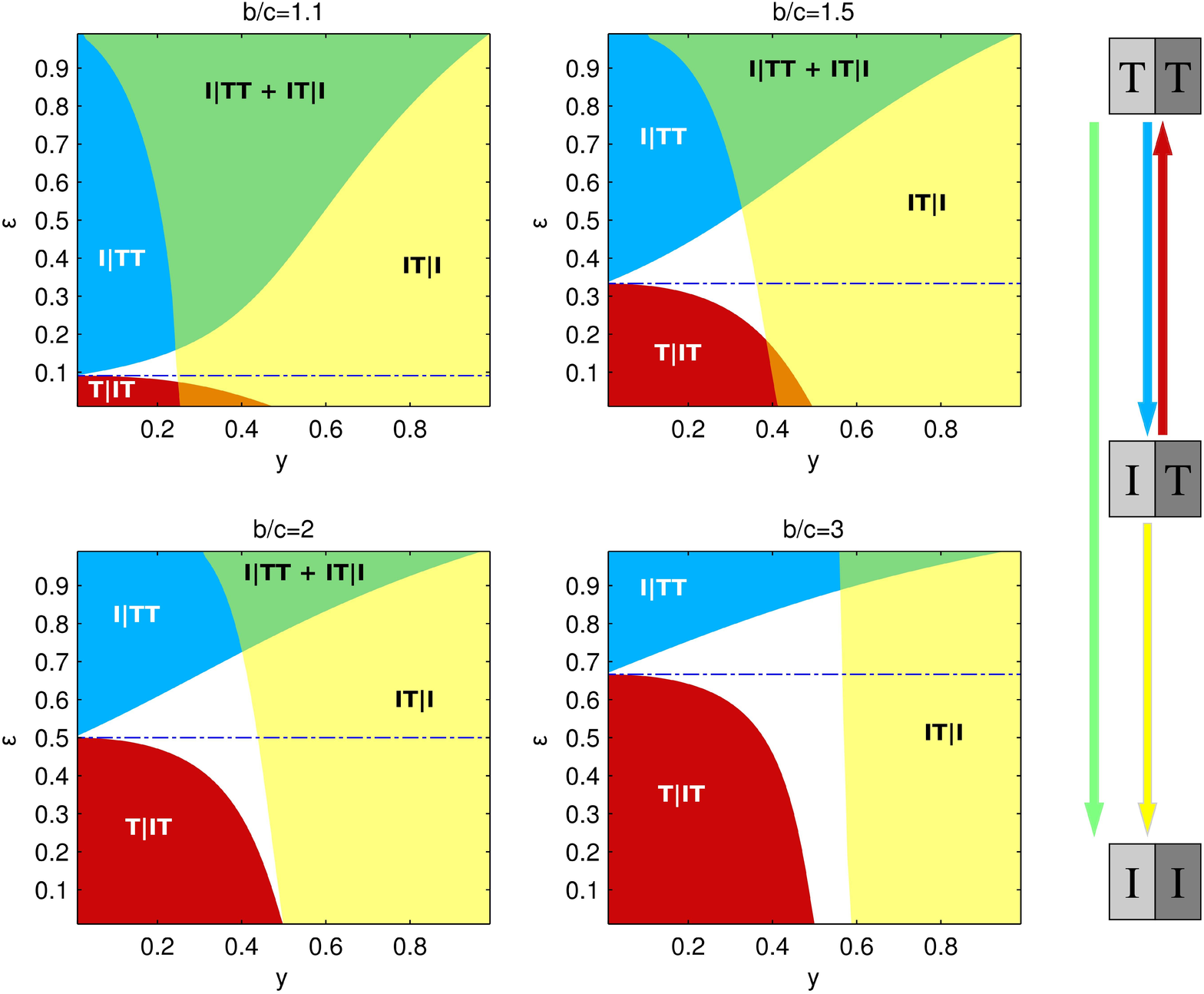}
\end{center}
\caption{
(Color online.)
Same as Fig.~\ref{fig:Ia} but for strategy IId.
}
\label{fig:IId}
\end{figure*}

\begin{figure*}
\begin{center}
\includegraphics[width=6in]{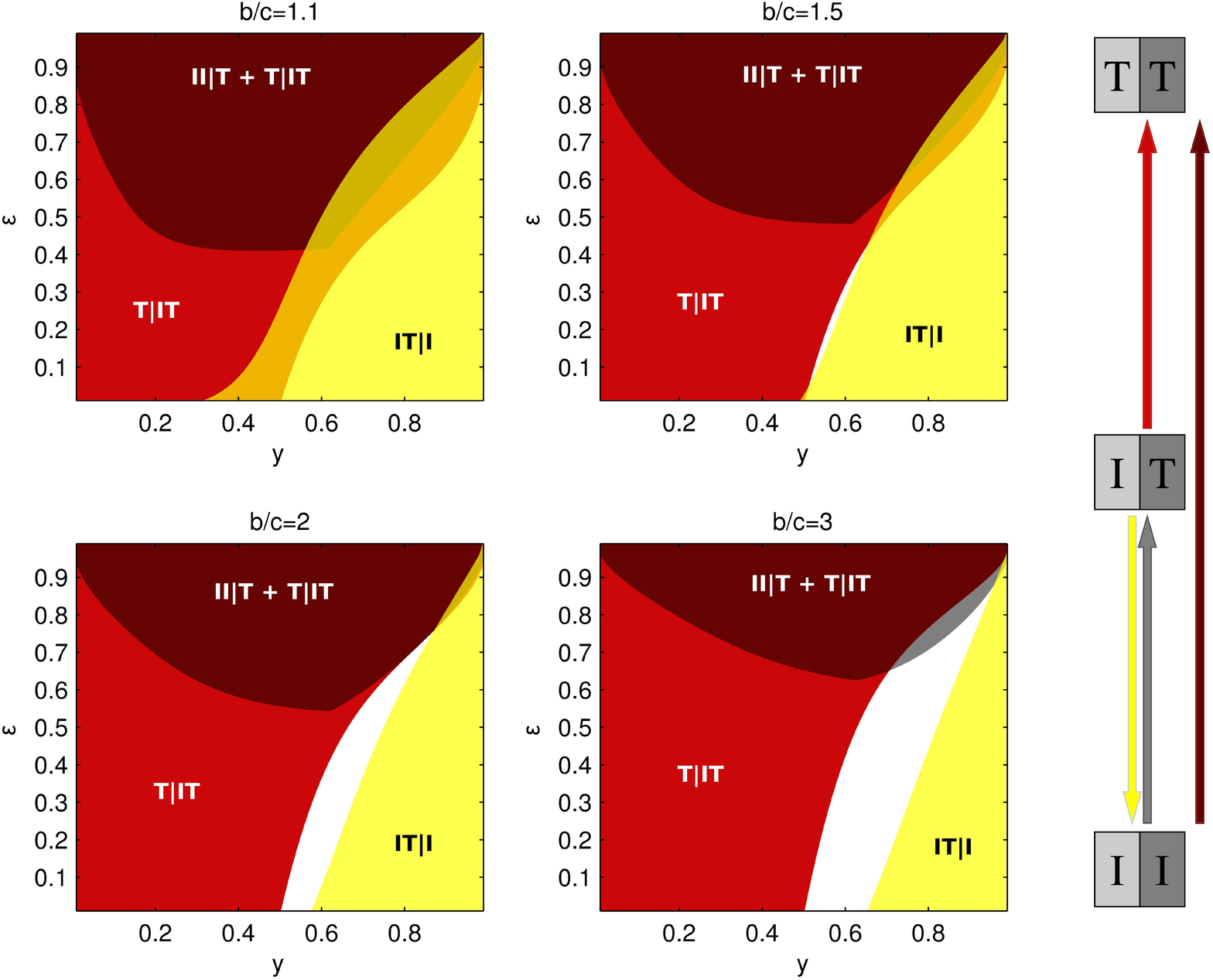}
\end{center}
\caption{
(Color online.)
Same as Fig.~\ref{fig:Ia} but for strategy IIIa, with the new invasion
scenario $II|T$.
}
\label{fig:IIIa}
\end{figure*}

\begin{figure*}
\begin{center}
\includegraphics[width=6in]{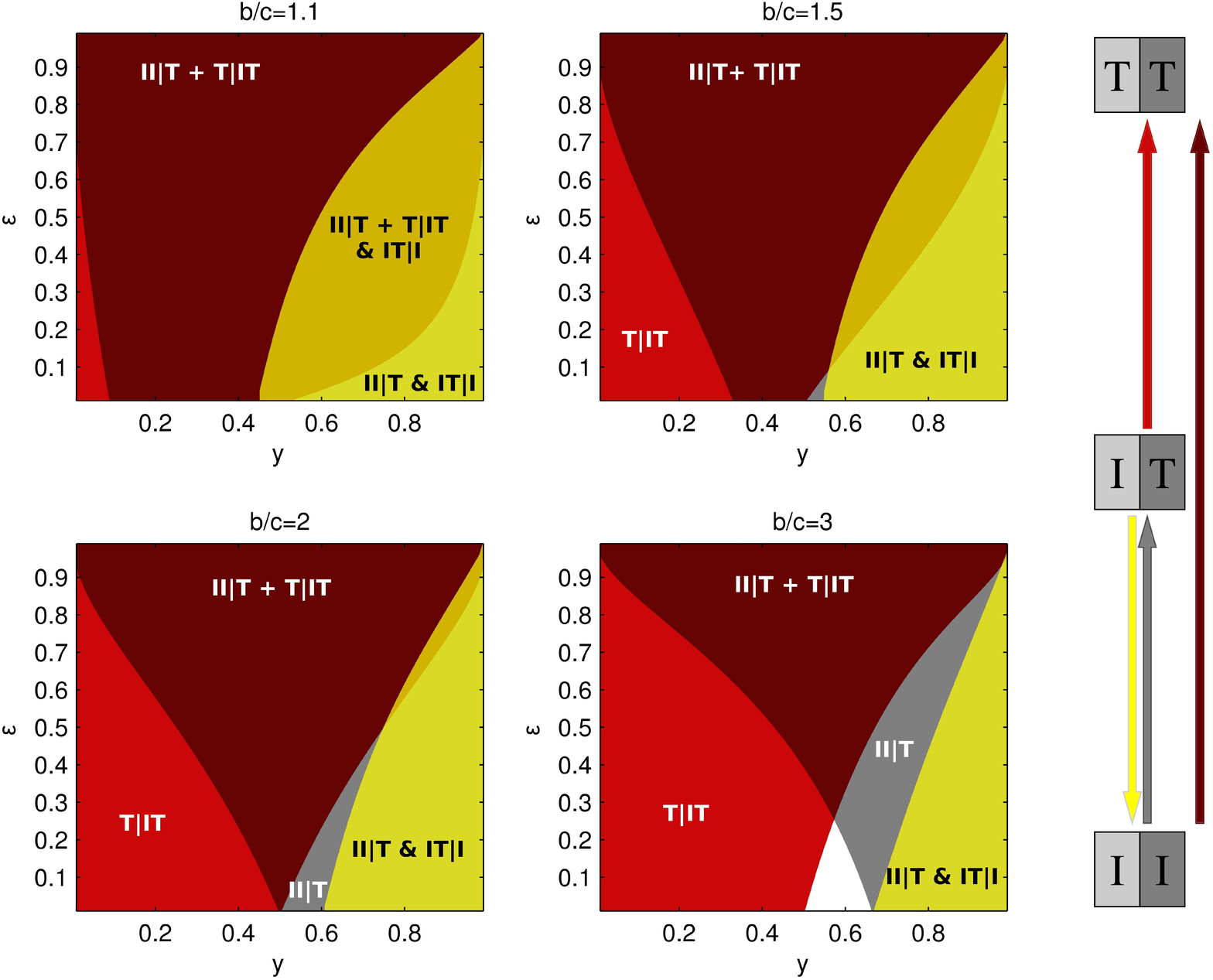}
\end{center}
\caption{
(Color online.)
Same as Fig.~\ref{fig:IIIa} but for strategy IIIb.
}
\label{fig:IIIb}
\end{figure*}

In what follows we will consider the different invasion scenarios for which
there exists a parameter region where invasion occurs. These regions are
determined numerically, and represented in Figs.~\ref{fig:Ia}--\ref{fig:IIIb}
as colored areas in a $\eA$--$y$ plane, where $y$ represents the fraction of
A-residents in the community. For the sake of clarity of the representations,
we will sometimes assume that the invader is of type A, and sometimes of type
B. Consistent with the previous section, we will denote
$t_M|t_At_B$---$t_X=\,$T,I being the tolerant/intolerant character of type X
individuals---whenever the mutant is of type A, and $t_At_B|t_M$ whenever the
mutant is of type B. (Notice, however, that $y$ will always represent the
fraction of A-residents, which will coincide with the type of mutants in the
former case, but not in the latter.) Changing the type of mutant will make
analyzing successive invasions $I|TT$ and $IT|I$ ---which would turn a fully
tolerant population into a fully intolerant one--- much easier.

In all invasion scenarios that we will
consider, we will just determine the stability of the resident population
against invasion by mutants, but not the final equilibrium (hence we do not
determine whether there is a turnover of the invaded population or a final
coexistence is reached).  Figures~\ref{fig:Ia}--\ref{fig:IId} correspond to
Group I and Group II strategies, whereas Figs.~\ref{fig:IIIa}--\ref{fig:IIIb}
correspond to Group III strategies.

The different scenarios are as
follows: $I|TT$, $T|IT$, $IT|I$, and $T|II$.
Overlaps of these regions are colored by a mixed color.

First of all, Figures~\ref{fig:Ia}--\ref{fig:IId} show that tolerant
communities that follow Group I and Group II strategies can be invaded by
intolerant individuals ($I|TT$) if $\eA$ is
sufficiently high and/or $b/c$ sufficiently low. Minorities are more prone to
undergo such a spread of intolerance. Second, if one of the two resident types
is intolerant and the other one is tolerant,  intolerant residents can be
invaded by tolerant mutants ($T|IT$) provided they are a minority
(in all cases that we have analysed $y<0.5$). However, this $T|IT$ region never
overlaps with the $I|TT$ one, so after a spread of intolerance the ``economic
conditions'' must change for tolerance to be restored. Third, if in a community
with tolerant and intolerant residents the latter are the majority, then this
strategy can spread also among the tolerant residents ($IT|I$),
thus dividing the community into two separate groups that dislike and never
help each other. And finally, a fully intolerant community resists invasion by
tolerant individuals for any combination of parameters. Hence, there is no way
to restore tolerance in a fully intolerant community by tuning the parameters
of the model (but see Section~\ref{sec:incentives}).

If we associate low $b/c$ ratios and high $\eA$ with strait conditions, the
interpretation emerging from these results is that economic stress favors the
spread of intolerance within minorities. Once it has invaded a group, tolerance
cannot be restored unless the economic conditions improve. And if intolerance
has eventually split the community, not even that can restore tolerance again.

Group III strategies are quite different
(Figs.~\ref{fig:IIIa}--\ref{fig:IIIb}). A tolerant population following any of
these strategies always resists the invasions of intolerant mutants regardless
of the parameter setting (i.e., invasion $I|TT$ never happens). Furthermore, if
one type is tolerant and the other one intolerant in the resident population,
tolerant mutants can invade the intolerant residents ($T|IT$).
Intolerant mutants can also invade the tolerant residents ($IT|I$),
but this occurs in a much narrower region of the diagram compared to
what happens for Group I and Group II strategies. As a matter of fact,
intolerant mutants only succeed if they try to invade a minority of tolerant
residents when the rest of the community is intolerant. Furthermore if we start
from a society where both subpopulations are intolerant, tolerant mutants have
a chance to invade the minority and spread their strategy ($II|T$)
---and perhaps eventually turn the community into tolerant individuals
in the brown region, overlap between $II|T$ and $T|IT$ invasion regions.
Overall, Group III strategies prove more stable against intolerance.

\subsection{Incentives to restore tolerance}
\label{sec:incentives}

\begin{figure*}
\begin{center}
\includegraphics[width=6in]{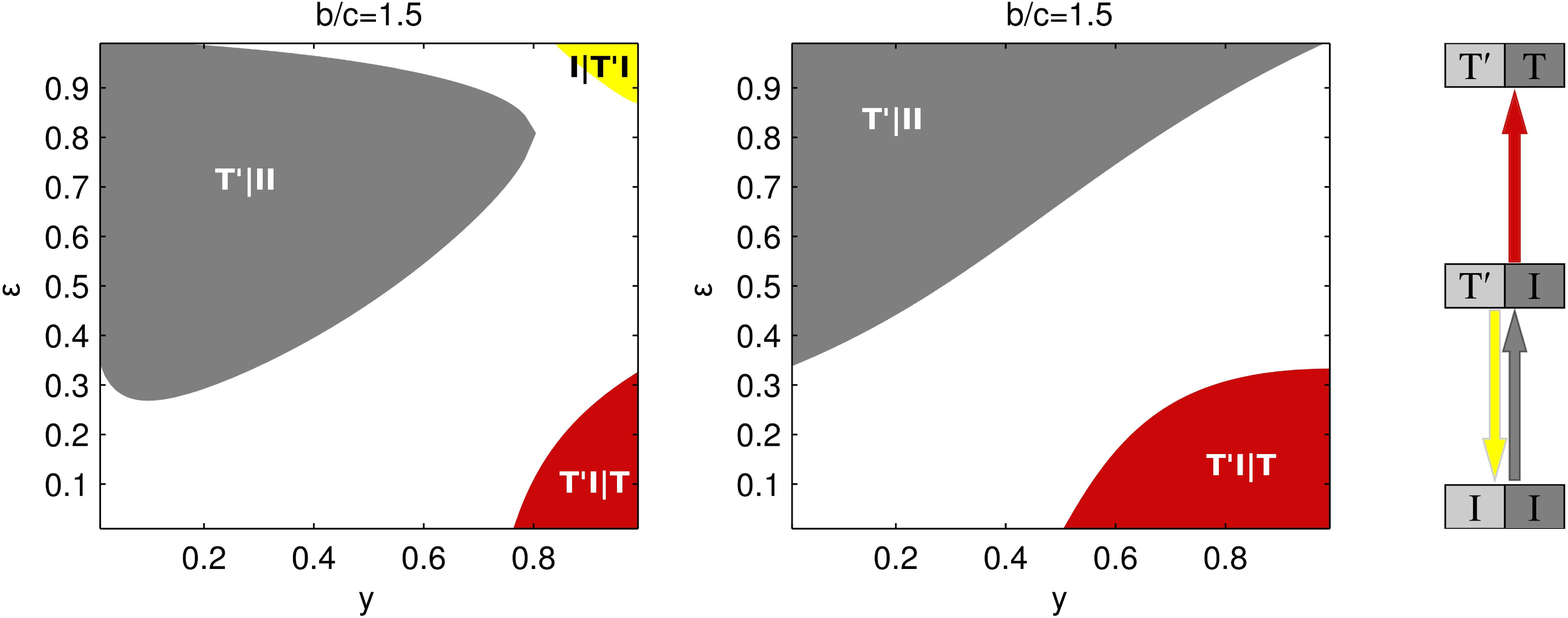}
\end{center}
\caption{(Color online.)
Conditions of $\eA$ and $y$ under which the invasions $T'|II$, $I|T'I$
 and $T'I|T$ are successful for Ib (left) and IId (right)
strategies. Incentivated cost $c'=0$.}
\label{fig:PD_no}
\end{figure*}

As shown in the previous section, if a tolerant community---or one of the two
subpopulations within it---following a Group I or a Group II strategy
eventually becomes intolerant, there is no way to restore tolerance again by
spreading it among the intolerant individuals. The only way to achieve this is by
``improving the economic conditions'' (i.e., decreasing $\eA$ and/or increasing
$b/c$). An alternative way to generate the same effect could be introducing an
exogenous incentive. We have implemented one such policy through a reduction of
the cost of helping others. In other words, we assume the presence of a
super-agent who partly subsidises the cost of helping individuals of a
subpopulation. This is done by lowering the cost of helping $B$-strategists.
By assuming this cost $c'<c$, Eqs.~\eqref{eq:WA1_gen} turn into
\begin{equation}
\begin{split}
W'(A|R)&=b\ \Theta_{R,A} -\Theta'_{A,R}, \\
W'(M|R)&=b\ \Theta_{R,M} -\Theta'_{M,R},
\end{split}
\label{eq:WA1_disc}
\end{equation}
where
\begin{equation}
\begin{split}
\Theta'_{A,R} &= c\ \theta_{A,A} +c'\ \theta_{A,B}, \\
\Theta'_{M,R} &= c\ \theta_{M,A} +c'\ \theta_{M,B}.
\end{split}
\end{equation}
The new scenarios where individuals of a class are given incentives to help the
other class are $I|T'T$, $T'|IT$, $I|T'I$, and $T'|II$, where primes mark
individuals who pay the lower cost $c'$ for helping the opposite class.
Equations~\eqref{eq:WA1_disc} must be used for these
individuals, whereas non-primed individuals follow Eqs.~\eqref{eq:WA1_gen},
as usual. Intolerant individuals are never marked since they never never help
the opposite class anyway.

\begin{figure*}
\begin{center}
\includegraphics[width=6in]{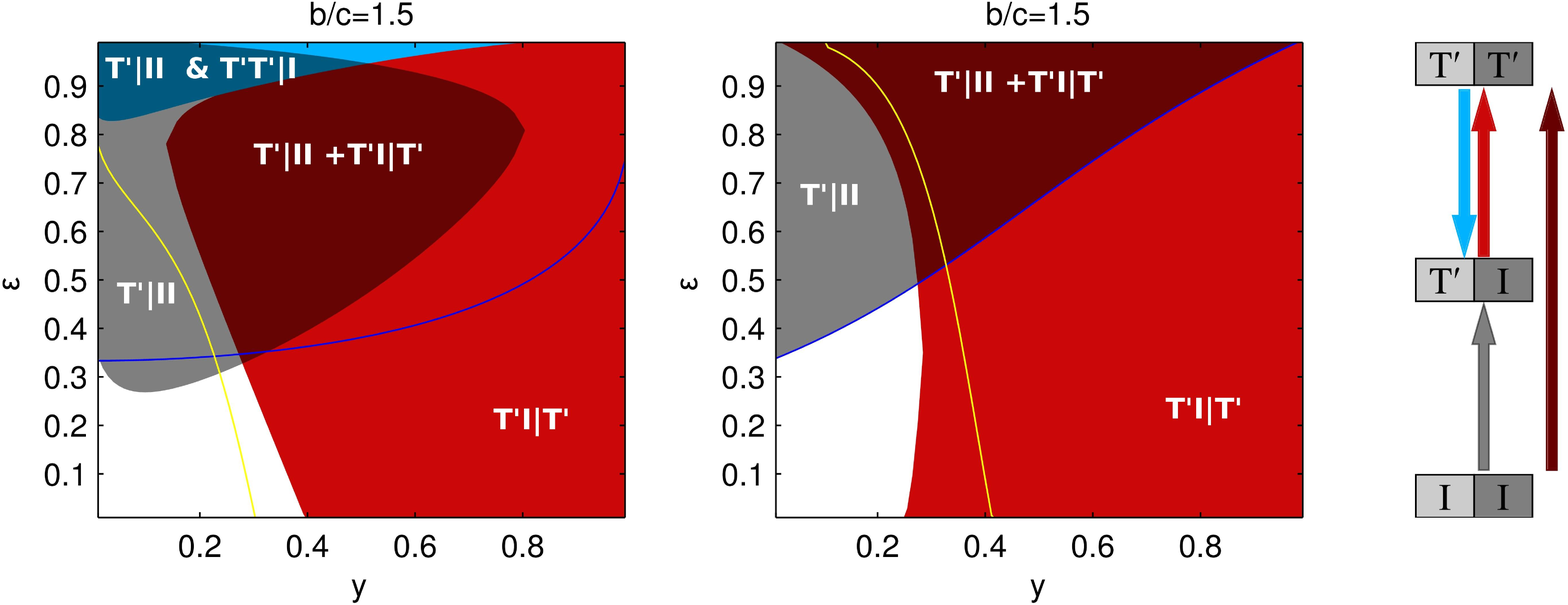}
\end{center}
\caption{(Color online.)
Conditions of $\eA$ and $y$ under which the invasions $T'|II$, $T'I|T'$
 and $T'T'|I$ are successful for Ib (left) and IId (right)
strategies. Dark (light) line shows the limit of the $I|TT$ ($IT|T$) region.
Incentivated cost $c'=0$.}
\label{fig:PD_yes}
\end{figure*}

From now on we focus only in Groups I and Group II strategies because tolerant
populations following Group III strategies are intolerance-proof.
The most effective incentive is reached for $c'=0$ (i.e., helping the opposite
class is free). Since strategies belonging to the same group have a very
similar behaviour, we show only one example for each group in
Figs.~\ref{fig:PD_no}-\ref{fig:PD_yes}.

Starting from a fully intolerant population, we represent in
Fig.~\ref{fig:PD_no} the region where incentives are able to promote tolerance
in one of the subpopulation ($T'|II$). Wherever this invasion
succeeds, the situation cannot be reverted, i.e., invasions $I|T'I$
never occur for the same parameter values ($I|T'I$ and regions $T'|II$ do
not overlap).  But on the other hand, intolerant residents in a $T'I$
population cannot be invaded by tolerant mutants ($T'I|T$) for the
same parameter values (again, $T'I|T$ and $T'|II$ regions do not overlap). In other
words, if incentives promote tolerance in one subpopulation of fully
intolerant individuals, the other subpopulation will still resist invasions by
tolerant mutants. Incentives are thus not enough to make a fully tolerant
community.

On the other hand, Figure~\ref{fig:PD_yes} shows that the region of
$T'I|T'$ invasions does overlap with the region of $T'|II$ invasions.
Thus, by providing incentives to both A and B groups, there are chances to
transform a fully intolerant community into a fully tolerant one. The latter is
also stable because the region where $T'T'|I$ invasions succeed does not
overlap with the $T'I|T'$ region.

Figure~\ref{fig:PD_yes} also shows that if an $II$ community has been formed
through a sequence of invasions $TT\rightarrow IT \rightarrow II$ from an
original $TT$ community, the values of $\eA$ and $y$ where this happens are
compatible with those for which full tolerance might be restored through
(extreme, i.e., $c'=0$) incentives.
The extreme case $c'=0$ is not very realistic, so in general one will have
$0<c'<c$, and the overlap regions are presumably smaller. But worse than that
is the fact that, as soon as incentives are removed, the scenarios of
Figs.~\ref{fig:Ia}--\ref{fig:IId} are restored and intolerance takes over
again. So the effect of incentives is not permanent.

\section{Discussion}
\label{sec:discussion}

We have studied under which conditions intolerance can invade a tolerant
community made of two distinguishable subgroups.
Our results show differences between two sets of the stable and coherent
strategies known in the literature of indirect reciprocity as leading eight
strategies. In the first set (Groups I and II) intolerance can invade a
tolerant subpopulation if the action error (which might associate to lack of
resources) is high enough and/or the benefit-to-cost ratio is low. This
invasion is more likely to occur if the invaded subpopulation is a minority.
Once one of the two subpopulation becomes intolerant, there is no way to
restore tolerance under the same economic conditions. Moreover, if the
intolerant minority is large enough, intolerance can also invade the tolerant
minority. A completely intolerant population cannot return to tolerance even if
economic conditions improve.

If one subpopulation is provided incentives to help the other, tolerance can be
restored for a limited region of parameters, but the other subpopulation does
not turn to tolerance by invasion. To transform a fully intolerant community
into a fully tolerant one both subpopulations must be stimulated to help the
other. Nonetheless, as soon as these incentives are removed, intolerance can
invade again. Thus, incentives have no permanent effect unless the economic
conditions are simultaneously improved. Incentives may provide a temporary mean
to achieve the goal, but improving the economic conditions is key to make the
effect perdurable.

In the second set of strategies (Group III), a fully tolerant population is
intolerance-proof. As a matter of fact, if we start off from a fully intolerant
community, tolerant mutants can invade it much easier than for the first set of
strategies (Groups I and II). The difference between the second set and the
first set---hence the reason why they resist intolerance---is that the only
way for an individual with bad reputation to achieve a good reputation is if
she helps to another individual with good reputation. In other words, these
strategies are less prone to forgive.

A way to interpret errors in actions is as a measure of global poverty of the
society. If resources are scarce, it may happen that an individual is forced
to deny help that would otherwise provide. But there may exist alternative
interpretations, e.g., it can also stand as the probability that an individual
is discovered when she secretly tries to cheat (defect). In this case, the
results suggest that intolerance spreads easier in a society with a high
proportion of known fraud. Note that in our model players do not know the
reasons that may drive someone to defect, i.e., they do not know if a player
defects on purpose or because she lacks the resources for it. The knowledge of
these inner motivations would involve much higher cognitive processes than the
simple direct observation that we are considering here.
  
In contrast to Nakamura and Masuda \cite{nakamura:2012}, we did not need to
make any mean-field approximation since reputations in our model come directly
from the observation of every interaction. The model we introduced is a
simplified description of very complex behaviours in real world. In particular,
our model considers that the whole society can be split into two classes whose
differentiation is based only on one set of traits. However many different sets
of traits can differentiate individuals in real life, and different people may
classify individuals according to different criteria. Even the relative
importance of some of these traits may change in time \cite{kurzban:2001}. In
complex scenarios with more classes of individuals and different ways to
classify them the results will presumably be different.

The study we have conducted has further limitations. Apart from the obvious
fact that using an indirect reciprocity model with binary reputation and
complete information certainly oversimplifies the problem (see
\cite{uchida:2010} and \cite{martinez:2013} for a discussion of these issues),
the main one is that we have considered only the onset of invasion by a small
fraction of mutants, but not the existence or absence of mixed equilibria (as
can be done, e.g., in direct reciprocity models \cite{martinez:2012}). In
other words, this study does not address the more difficult question whether
there is a turn over of the population or a coexistence of strategies. And
definitely simultaneous invasions by different type of mutants are excluded.
There is the implicit assumption that mutants emerge at a rate slower than the
time needed to reach an equilibrium. The existence of mixed equilibria thus
remains an open problem.

These limitations notwithstanding, our study provides some simple clues that
associate the emergence of intolerant attitudes with economic stress and the 
relative weight of the subgroups. Remarkably, given the simplicity of the
model, economic stress is recognised as one of the main predictors for the
emergence of intolerance \cite{gibson:2002}. As a matter of fact, this is a
typical prediction of social `scapegoat' theories \cite{tajfel:1981}, examples
of which are the research on the increased lynching of American blacks during
times of economic distress \cite{hovland:1940,dollard:1939} and the rise of
fascism and anti-Semitism in Germay during the interwar years
\cite{levin:1982,billing:1978}. Our study also makes some predictions that are
amenable to be tested in real situations, like the fact that intolerance
emerges predominantly within minorities, or---perhaps more importantly---that
societies with a stricter ethics are more resistant to intolerance than those
indulging a relaxed attitude towards defective behavior of their members. We
hope that this work stimulates empirical studies in these directions.

\section*{Acknowledgments}

This work is funded by the Ministerio de Econom\'{\i}a y Competitividad (Spain)
through grant PRODIEVO, and by the Comunidad de Madrid (Spain) through grant
MODELICO-CM. LAM-V was supported by a postdoctoral fellowship from Alianza 4
Universidades.

\appendix

\section{Homogeneous populations}
\label{sec:apen_aux}

It will prove useful to introduce the auxiliary function
\begin{equation}
U(x_1,x_2) =\sum_{\ag\bg} \chi_\ag(x_1) \chi_\bg(x_2) P_{\ag\bg},
\label{eq:fU}
\end{equation}
with $\chi_{\ag}(x)$ as defined by \eqref{eq:chi} and
\begin{equation}
P_{\ag\bg} = (1-\eA) m_{\ag\bg}(a_{\ag\bg}) + \eA m_{\ag\bg}(D)
\end{equation}
the probability that a donor  with reputation $\ag$ performing the action
$a_{\ag\bg}$ on a recipient with reputation $\bg$ is considered good according
to the moral rule $ m_{\ag\bg}(a)$.

In a homogeneous population---where all its members belong to same type---the
dynamics of the fraction of good individuals $g$ follows the differential
equation
\begin{equation}
 \frac{dg}{dt}=U(g,g)-g.
\end{equation}
In equilibrium $g=U(g,g)$, a quadratic equation with a unique stable
solution $0\leqslant g\leqslant 1$ \cite{martinez:2013}.

For later convenience, we will consider a continuous perturbation of the
above equation, namely
\begin{equation}
x=aU(x,x)+1-a,
\end{equation}
and denote $x=p(a)$ its stable solution in the interval $[0,1]$. Clearly
$p(1)=g$, the solution of the unperturbed equation.

Another function that we will use later is the solution of the linear equation
$x=aU(x,k)+1-a$, $0\leqslant a,k\leqslant 1$, which can be readily obtained as
\begin{equation}
\begin{split}
x &= \frac{1+a\lc kP_{01} + (1-k)P_{00}
-1\rc }{1-a\lc kP_{11}-(1-k)P_{10}+kP_{01} + (1-k)P_{00}\rc} \\
&\equiv \tilde{p}(k,a).
\end{split}
\end{equation}

\section{Equilibria for the different scenarios}
\label{sec:apen_equil}

We must obtain the equilibrium solutions of
Eqs.~\eqref{eq:evol_xA_gen}--\eqref{eq:evol_xM_gen} in order to compute the
probabilities $\theta_{i,j}$, which depend on $x_i^{\Lg_A\Lg_B\Lg_M}$. The set
of Eqs.~\eqref{eq:evol_xM_gen} is decoupled from
Eqs.~\eqref{eq:evol_xA_gen}--\eqref{eq:evol_xB2_gen}, hence $x_M^{\Lg_A*\Lg_M}$
can be calculated analytically from Eq.~\eqref{eq:evol_xM_gen} after solving 
Eqs.~\eqref{eq:evol_xA_gen}--\eqref{eq:evol_xB2_gen}. Thus one can readily
obtain $g_M^{M}$ and reduce Eq.~\eqref{eq:evol_xM_gen} to a linear system of
two equations in two unknowns (for example, $x_M^{G*G}$ and $x_M^{B*B}$). In
what follows, we describe the processes to complete our calculations for each
different scenario.

\subsection{$I|TT$ scenario}

This case considers the possible invasion of a fully tolerant resident
population by a small fraction of intolerant mutants. Since residents
do not distinguish classes, the resident population acts as a homogeneous
population. Therefore $x_i^{*\Lg_B\Lg_M}=x_i^{\Lg_A*\Lg_M}$,
$\theta_{B,A}=\theta_{A,B}=\theta_{A,A}$, $\theta_{B,M}=\theta_{A,M}$, and
$\theta_{M,B}=0$. The fraction of good residents is $g_B^{A}=x_A^{A}=p(1)$, as
previously discussed. According to the definition of $g_A^{A}$,
Eq.~\eqref{eq:g}, we have
\begin{align}
x_A^{G*B}&=g_A^{A}-x_A^{G*G},
\label{eq:x1GB} \\
x_A^{B*G}&=1-g_A^{A}-x_A^{B*B}.
\label{eq:xBG}
\end{align}
Then, the fraction $F^{\Lg_A\Lg_M}_{A,B}$ introduced in Eq.~\eqref{eq:evol_xA_gen} can be expressed in
this scenario as
\begin{align}
 F^{\Lg_A\Lg_M}_{A,B}=&\sum_{\ag\bg} \chi_{\ag}(g^A_A) \chi_{\bg}(g^A_A)
P_A^{\Lg_A\Lg_M}(\ag\bg), \\
\begin{split}
P_A^{\Lg_A\Lg_M}(\ag\bg)\ =& \ (1-\eA)\,  \delta  \lp
\Lg_A, m_{\ag\bg}(a_{\ag\bg}) \rp \\
&\times \delta \lp \Lg_M,  1-a_{\ag\bg}\rp \\
&+ \eA\, \delta \lp \Lg_A, m_{\ag\bg}(D)\rp \times \delta \lp \Lg_M, 1 \rp.
\end{split}
\end{align}

Once we numerically solve the two Eqs.~\eqref{eq:evol_xA_gen},
Eq.~\eqref{eq:evol_xM_gen}, which corresponds to the intolerant mutants, can be
solved analytically by taking into account that
\begin{align}
F^{\Lg_A\Lg_M}_{M,B}&=\sum_{\ag\bg} \chi_{\ag}(g^A_M) \chi_{\bg}(g^A_B)
P_M^{\Lg_A\Lg_M}(\ag\bg), \\ P_M^{\Lg_A\Lg_M}(\ag\bg)\ &=\  \delta \lp \Lg_A,
m_{\ag\bg}(D) \rp \delta \lp \Lg_M, 1 \rp.
\end{align}
If at equilibrium, Eq.~\eqref{eq:evol_xM_gen} is summed over $\Lg_A$, we find
that $g_M^{M}$ is given by $\tilde p(g_A^{M},y)$. Then the set of
Eqs.~\eqref{eq:evol_xM_gen} simplifies to just two equations.

\subsection{$T|IT$ scenario}

In this scenario a small fraction of tolerant individuals tries to invade a
population where individuals of their own class are intolerant
($\theta_{A,B}=0$), whereas those of the other class are tolerant. Since mutants
and B-residents are both tolerant, they will judge every player equally.
Then $x_i^{\Lg_A\Lg_B*}=x_i^{\Lg_A*\Lg_M}$.

The dynamics of the intolerant A-strategists is described by
Eq.~\eqref{eq:evol_xA_gen}, where
\begin{align}
F^{\Lg_A\Lg_M}_{A,B}&=\sum_{\ag\bg} \chi_{\ag}(g^M_A) \chi_{\bg}(g^M_B)
P_A^{\Lg_A\Lg_M}(\ag\bg), \\ P_A^{\Lg_A\Lg_M}(\ag\bg)\ &=\  \delta \lp \Lg_M,
m_{\ag\bg}(D) \rp \delta \lp \Lg_A, 1 \rp.
\end{align}
Summing Eqs.~\eqref{eq:evol_xA_gen} over $\Lg_M$ at equilibrium we obtain
$g_A^{A}=p(y)$. In this scenario we also need to calculate $g_B^{M}$.
Since $B$-mutant's judgements are the same as those of B-residents,
$g_B^{M}$ is given at equilibrium by
\begin{equation}
g_B^{M}\ =\ (1-y) U(g_B^{M},g_B^{M}) + y U(g_B^{M},g_M^{A}).
\label{eq:equil_xB_TIT}
\end{equation}
Therefore, we need to solve a system of three equations (two from
Eqs.~\eqref{eq:evol_xA_gen} plus Eq.~\eqref{eq:equil_xB_TIT}), choosing,
for instance, $x_A^{G*G}$, $x_A^{B*B}$, and $g_B^{M}$ as unknowns.
After solving this system numerically, the set of Eqs.~\eqref{eq:evol_xM_gen},
where
\begin{align}
 F^{\Lg_A\Lg_M}_{M,B}=&\sum_{\ag\bg} \chi_{\ag}(g^M_M)
 \chi_{\bg}(g^B_M) P_M^{\Lg_A\Lg_M}(\ag\bg)\ \\
\begin{split}
 P_M^{\Lg_A\Lg_M}(\ag\bg)\ =&\ (1-\eA)\, \delta \lp
\Lg_M, m_{\ag\bg}(a_{\ag\bg}) \rp \\
&\times \delta \lp \Lg_A,  1-a_{\ag\bg}\rp \\
&+ \eA\, \delta \lp \Lg_M, m_{\ag\bg}(D)\rp \times \delta \lp \Lg_A, 1 \rp,
\end{split}
\end{align}
can be solved analytically. For that we take into account that summing
over $\Lg_A$ at equilibrium yields $g_M^{M}=\tilde p\lp y g_A^{M}+(1-y)
g_B^{M} , 1 \rp$.

\subsection{$I|TI$ scenario}

In this scenario a small fraction of intolerant mutants tries to invade a
population where individuals of their own class are tolerant whereas those of
the other class are intolerant. Hence
$\theta_{B,A}=\theta_{B,M}=\theta_{M,B}=0$.

In order to calculate $g_B^A$ we first need to compute $x_B^{\Lg_A\Lg_B*}$
through Eq.~\eqref{eq:evol_xB1_gen}, where
\begin{align}
F^{\Lg_A\Lg_B}_{B,A}&=\sum_{\ag\bg} \chi_{\ag}(g^A_B) \chi_{\bg}(g^A_A)
P_B^{\Lg_A\Lg_B}(\ag\bg), \\ P_B^{\Lg_A\Lg_B}(\ag\bg)\ &=\  \delta \lp \Lg_A,
m_{\ag\bg}(D) \rp \delta \lp \Lg_B, 1 \rp,
\end{align}
and summing over $\Lg_A$ at equilibrium one obtains that $g_B^{B}= p(1-y)$.
This way, we reduce the set of Eqs.~\eqref{eq:evol_xB1_gen} to just two
equations.

Now, the dynamics of the tolerant individuals is given by
Eq.~\eqref{eq:evol_xA_gen}, with
\begin{align}
 F^{\Lg_A\Lg_M}_{A,B}=&\sum_{\ag\bg} \chi_{\ag}(g^A_A) \chi_{\bg}(g^A_B)
P_A^{\Lg_A\Lg_M}(\ag\bg), \\
\begin{split}
P_A^{\Lg_A\Lg_M}(\ag\bg)\ =&\ (1-\eA)\, \delta  \lp
\Lg_A, m_{\ag\bg}(a_{\ag\bg}) \rp \\
&\times \delta \lp \Lg_M,  1-a_{\ag\bg}\rp \\
&+ \eA\, \delta \lp \Lg_A, m_{\ag\bg}(D)\rp \times \delta \lp \Lg_M, 1 \rp.
\end{split}
\end{align}
The only simplification that works in this case is to take into account
that $x_A^{B*G}=1-x_A^{G*G}-x_A^{G*B}-x_A^{B*B}$. Thus we need to solve
numerically five coupled equations: three from Eqs.~\eqref{eq:evol_xA_gen} and
two from Eqs.~\eqref{eq:evol_xB1_gen}, corresponding to the unknowns
$x_B^{GG*}$, $x_B^{BB*}$, $x_A^{G*G}$, $x_B^{G*B}$, and $x_B^{B*B}$.

The dynamics of the mutants is solved from Eq.~\eqref{eq:evol_xM_gen}, where
\begin{align}
F^{\Lg_A\Lg_M}_{M,B}&=\sum_{\ag\bg} \chi_{\ag}(g^A_M) \chi_{\bg}(g^A_B)
P_M^{\Lg_A\Lg_M}(\ag\bg), \\ P_M^{\Lg_A\Lg_M}(\ag\bg)\ &=\  \delta \lp \Lg_A,
m_{\ag\bg}(D) \rp \ \delta \lp \Lg_M, 1 \rp.
\end{align}
Summing over $\Lg_A$ at equilibrium yields $g_M^{M}=\tilde p(g_A^{M},y)$, and
transforms the last equation to just two linear equations.

\subsection{$T|II$ scenario}

In this last scenario a small fraction of tolerant mutants tries to invade a
completely intolerant population. Then
$\theta_{A,B}=\theta_{B,A}=\theta_{B,M}=0$.

The dynamics for the intolerant A-residents is described by
Eq.~\eqref{eq:evol_xA_gen} with
\begin{align}
F^{\Lg_A\Lg_M}_{A,B}&=\sum_{\ag\bg} \chi_{\ag}(g^M_A) \chi_{\bg}(g^M_B)
P_A^{\Lg_A\Lg_M}(\ag\bg), \\ P_A^{\Lg_A\Lg_M}(\ag\bg)\ &=\  \delta \lp \Lg_M,
m_{\ag\bg}(D) \rp \delta \lp \Lg_A, 1 \rp.
\end{align}
Summing over $\Lg_M$ at equilibrium yields $g_A^{A} = p(y)$.

In order to calculate $g_B^{M}$, we need first to compute $x_B^{*\Lg_B\Lg_M}$.
The dynamics for these fractions is described by Eq.~\eqref{eq:evol_xB2_gen},
where
\begin{align}
F^{\Lg_B\Lg_M}_{B,A}&=\sum_{\ag\bg} \chi_{\ag}(g^M_B) \chi_{\bg}(g^M_A)
P_B^{\Lg_B\Lg_M}(\ag\bg), \\ P_B^{\Lg_B\Lg_M}(\ag\bg)\ &=\  \delta \lp \Lg_M,
m_{\ag\bg}(D) \rp \delta \lp \Lg_B, 1 \rp,
\end{align}
and summing again over $\Lg_M$ we obtain that $g_B^{B} = p(1-y)$. Thus we have
four coupled equations (two from Eqs.~\eqref{eq:evol_xB1_gen} and two from
Eq.~\eqref{eq:evol_xB2_gen}) for the unknowns $x_A^{G*G}$, $x_A^{B*B}$,
$x_B^{*GG}$, and $x_B^{*BB}$, which need to be solved numerically.

On the other hand, the dynamics of the tolerant mutants is given by
Eq.~\eqref{eq:evol_xB2_gen}, where
\begin{align}
 F^{\Lg_A\Lg_M}_{M,B}=&\sum_{\ag\bg} \chi_{\ag}(g^M_M) \chi_{\bg}(g^M_B)
P_M^{\Lg_A\Lg_M}(\ag\bg), \\ 
\begin{split}
P_M^{\Lg_A\Lg_M}(\ag\bg)\ =&\ (1-\eA)\, \delta  \lp
\Lg_M, m_{\ag\bg}(a_{\ag\bg}) \rp \\
&\times \delta \lp \Lg_A, 1-a_{\ag\bg}\rp \\ 
&+ \eA\, \delta \lp \Lg_M, m_{\ag\bg}(D)\rp \delta \lp \Lg_A, 1 \rp.
\end{split}
\end{align}
This set of equations can be reduced to just two because summing over $\Lg_A$ at
equilibrium yields $g_M^{M}=\tilde p( yg_A^{M}+(1-y)g_B^{M},1)$.

\bibliographystyle{apsrev4-1}
\bibliography{evol-coop}

\end{document}